\documentclass[aps, prb, showpacs, superscriptaddress ,twocolumn ]{revtex4}
\usepackage{amsmath}
\usepackage{graphicx}
\usepackage{sidecap}
\usepackage{hyperref}
\usepackage{ulem}

\begin{document}

\title{Rotation symmetry breaking in La$_{2-x}$Sr$_{x}$CuO$_4$ revealed by ARPES}

\author{E. Razzoli}
\altaffiliation[Present address:]{ Quantum Matter Institute, Department of Physics and Astronomy, University of British Columbia, Vancouver, British Columbia V6T 1Z1, Canada}
\affiliation{Swiss Light Source, Paul Scherrer Institute, CH-5232 Villigen PSI, Switzerland}
\affiliation{D{\'e}partement de Physique and Fribourg Center for Nanomaterials, Universit\'e de Fribourg, CH-1700 Fribourg, Switzerland}

\author{C. E. Matt}
\affiliation{Swiss Light Source, Paul Scherrer Institute, CH-5232 Villigen PSI, Switzerland}
\affiliation{Laboratory for Solid State Physics, ETH Zurich, CH-8093 Zurich, Switzerland}

\author{Y. Sassa}
\affiliation{Swiss Light Source, Paul Scherrer Institute, CH-5232 Villigen PSI, Switzerland}
\affiliation{Laboratory for Solid State Physics, ETH Zurich, CH-8093 Zurich, Switzerland}
\affiliation{Department of Physics and Astronomy, Uppsala University, S-75121 Uppsala, Sweden}

\author{M. M\aa{}nsson}
\affiliation{Laboratory for Quantum Magnetism (LQM),Ecole Polytechnique F\'{e}d\'{e}rale de Lausanne (EPFL), Station 3, CH-1015 Lausanne, Switzerland}
\affiliation{Laboratory for Neutron Scattering, Paul Scherrer Institut, CH-5232 Villigen PSI, Switzerland}
\affiliation{KTH Royal Institute of Technology, Materials Physics, Electrum 229, 164 40 Kista, Stockholm, Sweden}

\author{O. Tjernberg}
\affiliation{KTH Royal Institute of Technology, Materials Physics, Electrum 229, 164 40 Kista, Stockholm, Sweden}

\author{G. Drachuck}
\affiliation{Physics Department, Technion, Israel Institute of Technology, Haifa 32000, Israel}

\author{M. Monomo}
\affiliation{Department of Applied Sciences, Muroran Institute of Technology, Muroran 050-8585, Japan}

\author{M. Oda}
\affiliation{Department of Physics, Hokkaido University, Sapporo 060-0810, Japan}

\author{T. Kurosawa}
\affiliation{Department of Physics, Hokkaido University, Sapporo 060-0810, Japan}

\author{Y. Huang}
\affiliation{Beijing National Laboratory for Condensed Matter Physics, and Institute of Physics, Chinese Academy of Sciences, Beijing 100190, China}

\author{N. C. Plumb}
\affiliation{Swiss Light Source, Paul Scherrer Institute, CH-5232 Villigen PSI, Switzerland}

\author{M. Radovic}
\affiliation{Swiss Light Source, Paul Scherrer Institute, CH-5232 Villigen PSI, Switzerland}

\author{A. Keren}
\affiliation{Physics Department, Technion, Israel Institute of Technology, Haifa 32000, Israel}

\author{L. Patthey}
\affiliation{Swiss Light Source, Paul Scherrer Institute, CH-5232 Villigen PSI, Switzerland}

\author{J. Mesot}
\affiliation{Swiss Light Source, Paul Scherrer Institute, CH-5232 Villigen PSI, Switzerland}
\affiliation{Laboratory for Solid State Physics, ETH Zurich, CH-8093 Zurich, Switzerland}
\affiliation{Institut de la Matiere Complexe, EPF Lausanne, CH-1015, Lausanne, Switzerland}

\author{M. Shi}
\affiliation{Swiss Light Source, Paul Scherrer Institute, CH-5232 Villigen PSI, Switzerland}


\begin{abstract}

Using angle-resolved photoemission spectroscopy it is revealed that in the vicinity of optimal doping the electronic structure of La$_{2-x}$Sr$_{x}$CuO$_4$ cuprate undergoes an electronic reconstruction associated with a wave vector $\boldsymbol{q}_{\boldsymbol{a}}=(\pi, 0)$. The reconstructed Fermi surface and folded band are distinct to the shadow bands observed in BSCCO cuprates and in underdoped La$_{2-x}$Sr$_{x}$CuO$_4$ with $x \le 0.12$, which shift the primary band along the zone diagonal direction. Furthermore the folded bands appear only with $\boldsymbol{q}_{\boldsymbol{a}}=(\pi, 0)$ vector, but not with $\boldsymbol{q}_{\boldsymbol{b}}= (0, \pi)$. We demonstrate that the absence of $\boldsymbol{q}_{\boldsymbol{b}}$ reconstruction is not due to the matrix-element effects in the photoemission process, which indicates the four-fold symmetry is broken in the system.

\end{abstract}

\date{\today}
\maketitle

\section{Introduction}

Since the discovery of high-temperature cuprate superconductors, the study of various instabilities (e.g. magnetic and charge order) emerging in close proximity to superconductivity has attracted much attention. Significant efforts have been devoted to reveal other instabilities than superconducting one, as a function of doping and temperature, and how they are intertwined with the superconductivity. Recently an incipient incommensurate charge-density-wave (CDW) in underdoped YBa$_2$Cu$_3$O$_{6+y}$ (YBCO) with hole concentrations in the range of 0.09 to 0.13 per planar Cu ion has been reported independently from high-energy X-ray diffraction \cite{Chang2012} and resonant soft X-ray scattering \cite{Ghiringhelli2012} experiments. Similar charge modulations along the Cu-O bonding directions of underdoped Bi$_2$Sr$_{2-x}$La$_x$CuO$_{6+\delta}$ (Bi2201) and Bi$_2$Sr$_2$CaCu$_2$O$_{8+\delta}$ (Bi2212) have also been observed in scanning-tunneling microscopy and resonant x-ray scattering measurements, with the ordering vector approaching to a commensurate wave vector $(0.25 \times \pi /a)$ when the hole doping is increasing \cite{Comin2014, SilvaNeto2014}. For underdoped La-based ``214'' family of cuprates (La$_{2-x-y}$(Sr,Ba)$_x$(Nd,Eu)$_y$CuO$_4$) at a doping level $x  \approx 1/8$ an unidirectional modulated antiferromagnetism \cite{Plate2005} combined with a commensurate charge modulation of period 4 lattice constant (stripe order) has long been identified, and at the same doping level the superconducting transition temperature is dramatically reduced \cite{Tranquada1995, Tranquada1996}.  The presence of stripe order  in   La$_{2-x}$Sr$_x$CuO$_4$ has been debated for long time and only recently scattering measurements have shown evidence for a CDW with a  wave vector $\boldsymbol{q}$ displaying a  doping dependence similar to the one observed in Bi2201 and La$_{2-x}$Ba$_x$CuO$_4$ \cite{Croft2014, Christensen2014, Thampy2014}. However, so far, angle-resolved photoemission spectroscopy (ARPES) measurements have not shown any evidence of band folding associated with charge ordering along the Cu-O bond direction in cuprates \cite{Kohsaka2008, Meng2009, Kanigel2006}.

In this letter, applying ARPES to nearly optimally doped LSCO ($x = 0.15, 0.17$) we show that in the superconducting phase the Fermi surface (FS) is reconstructed along the Cu-O bond direction associated with a wave vector $\boldsymbol{q}_{\boldsymbol{a}} = (\pi,0)$. This wave vector could be related to the second harmonic of an incipient CDW in the region of optimal doping, which is the smooth continuation with doping of the incipient CDW  as observed in other cuprates, or could point to a new instability in the system.

%
%
%
\begin{figure}
\includegraphics[width=0.5\textwidth]{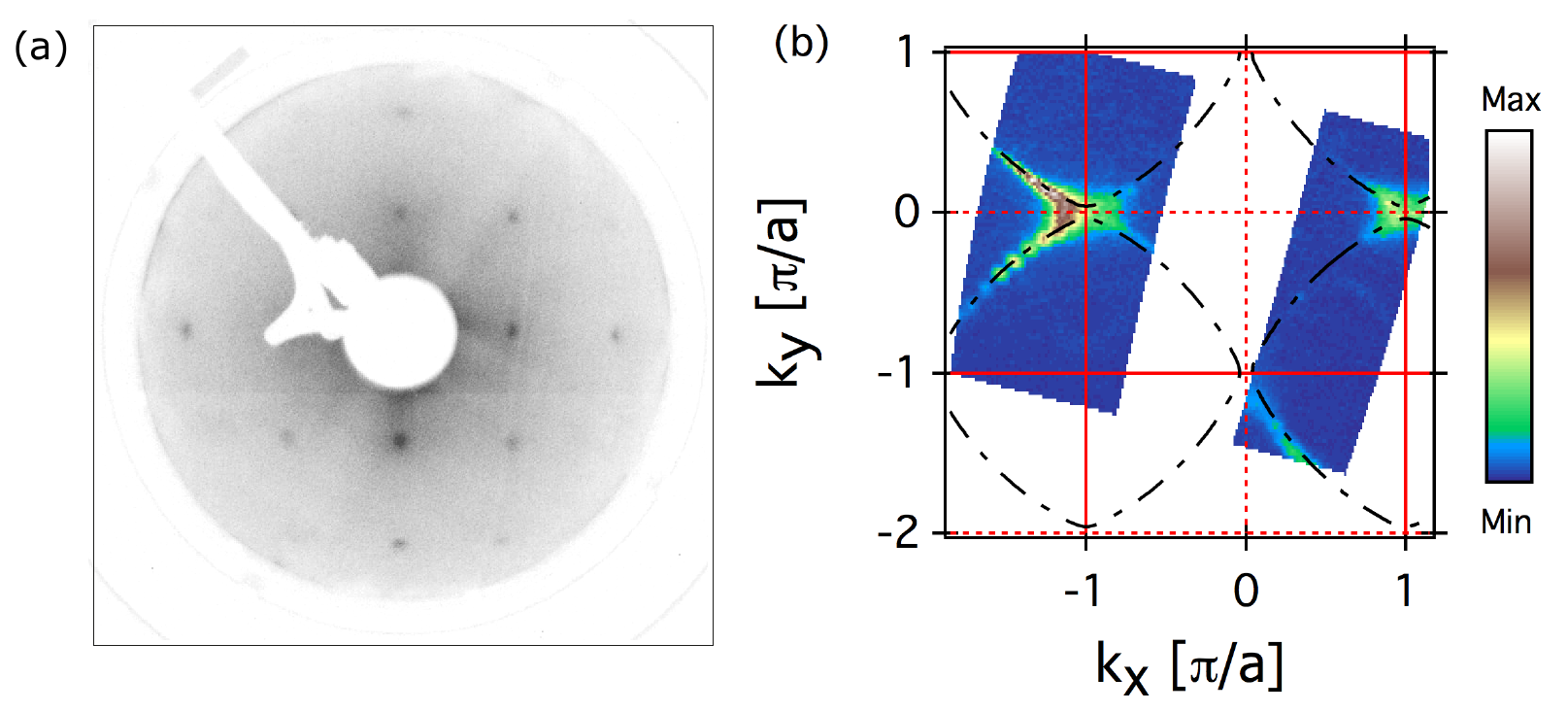}
\caption{(a) LEED pattern taken at $T= 12$ K and $E_e=185$ eV.  (b) FS mapping for LSCO $x=0.17$. Dashed-pointed black line is a TB fit as explained in the text.}\label{FS_Map_and_LEED}
\end{figure}
%
%
%
\begin{figure*}[ht]
\includegraphics[width=0.9\textwidth]{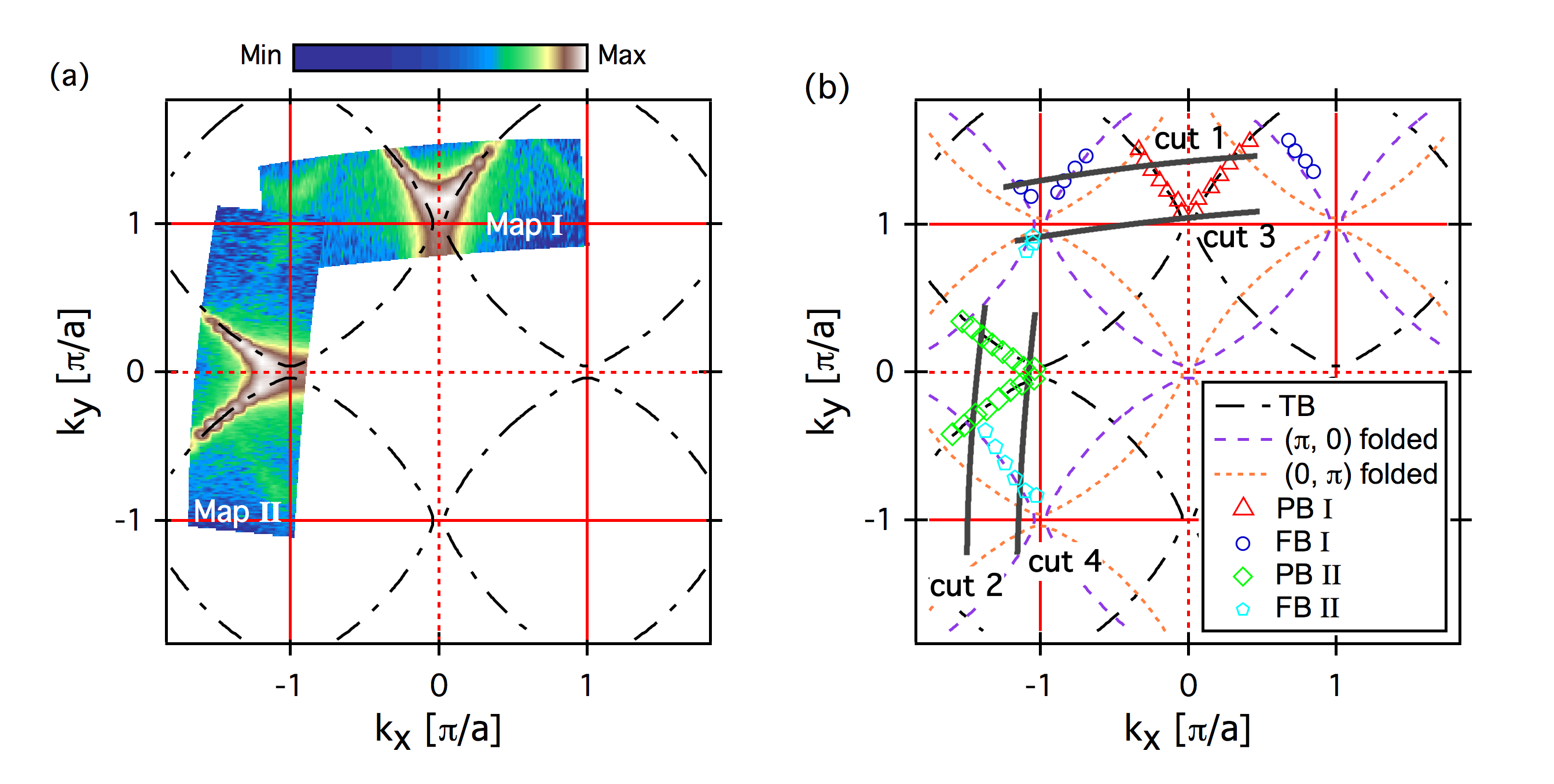}
\caption{  ARPES Intensity plots of La$_{2-x}$Sr$_x$CuO$_4$ ($x = 0.17$). The data were taken at 12 K. (a) FS intensity maps in the $k_x - k_y$ plane at $h\nu= 55$ eV and $T=12$ K . Map II is the same as Map I but after a 90$^\circ$ rotation of the sample. 
The FS maps  in I-II are obtained by integrating ARPES spectral weight in an energy window of $E_F \pm 20$ meV. (b)  Red  triangles (green  rombus) and blue  circles (light-blue pentagons) are the $k_F$ for the primary band (PB) and the folded band (FB) extracted from the MDCs peaks for the sample oriented like in Map I (Map II), respectively.  Dashed-pointed black line is a TB fit to the PB data, dashed violet line is the TB FS shifted by $\boldsymbol{q}_{\boldsymbol{a}} = (\pi,0)$, and dotted orange line is the TB FS shifted by $\boldsymbol{q}_{\boldsymbol{b}} = (0, \pi)$ . }\label{FS_folding17}
\end{figure*}

\section{Experimental details}
ARPES experiments were carried out at the Surface and Interface Spectroscopy beamline at the Swiss Light Source of Paul Sherrer intitute on single crystals La$_{2-x}$Sr$_x$CuO$_4$ (LSCO). The doping values used for the measurements are $x = 0.15$ and 0.17 with superconducting transition temperature ($T_c$) of 38 K and 35 K, respectively. The crystals were grown in traveling solvent floating zone furnaces. All samples were characterized by x-ray diffraction, and their superconducting transitions were determined by magnetization measurements. Circularly polarized light with $h\nu = 55$ eV was used in order to maximize the signal. The spectra were recorded with Scienta  R4000 analyzers. The energy and angle resolutions were $\sim 15$ meV and 0.1 - 0.15$^\circ$, respectively. The Fermi level was determined by recording photoemission spectra from polycrystalline copper on the sample holder. The samples were cleaved in situ by using a specially designed cleaver \cite{Mansson2007}. Low-energy electron diffraction analysis of the cleaved samples shows a clear ($1 \times  1$) pattern with no sign of surface reconstruction [see Fig. \ref{FS_Map_and_LEED}(a)]. During the measurements, the base pressure always remained less than $5 \times 10^{-11}$ mbar.

\section{Experimental results}

Figure \ref{FS_Map_and_LEED}(b) shows the spectral weight mapping in $k$-space at Fermi level ($E_F$). The superimposed dashed black line is the FS obtained from a tight binding (TB) fit to the $k_F$ extracted from the peak positions of momentum distribution curves (MDC) at $E_F$ and to the MDC peak positions, as a function of binding energy, along the zone diagonal (nodal dispersion).  The basis functions and the obtained fitting coefficients with the constraint $t_3/t_2 = -1/2$ \cite{Pavarini2001} are listed in Table \ref{Table_TB}. Luttinger sum rule \cite{Luttinger1960} gives a hole concentration of $x\sim0.19$, slightly bigger than the nominal doping $x=0.17$, in agreement with the observation in early studies \cite{Yoshida2006}.

\begin{table}
\caption{Tight-binding coefficient and basis functions used in fitting the experimental data. The second column lists the coefficient of each term (meV) following the convention: $\epsilon(\boldsymbol{k}) = \sum t_i \eta_i(\boldsymbol{k})$.}\label{Table_TB} 
\begin{ruledtabular}
\begin{tabular}{@{\qquad}c c c@{\qquad}}
$i$ &  $t_i$	& $\eta_i(\boldsymbol{k})$\\
\hline
0	&	134	& 1\\
1	&	181	& -2[$\cos(k_xa)+\cos(k_ya)$]\\
2	&   -23	& -4[$\cos(k_xa)\cos(k_ya)$]\\
3	&	 12	& -2[$\cos(2k_xa)+\cos(2k_ya)$]\\
\end{tabular}
\end{ruledtabular}
\end{table}

\begin{figure}
\includegraphics[width=0.49\textwidth]{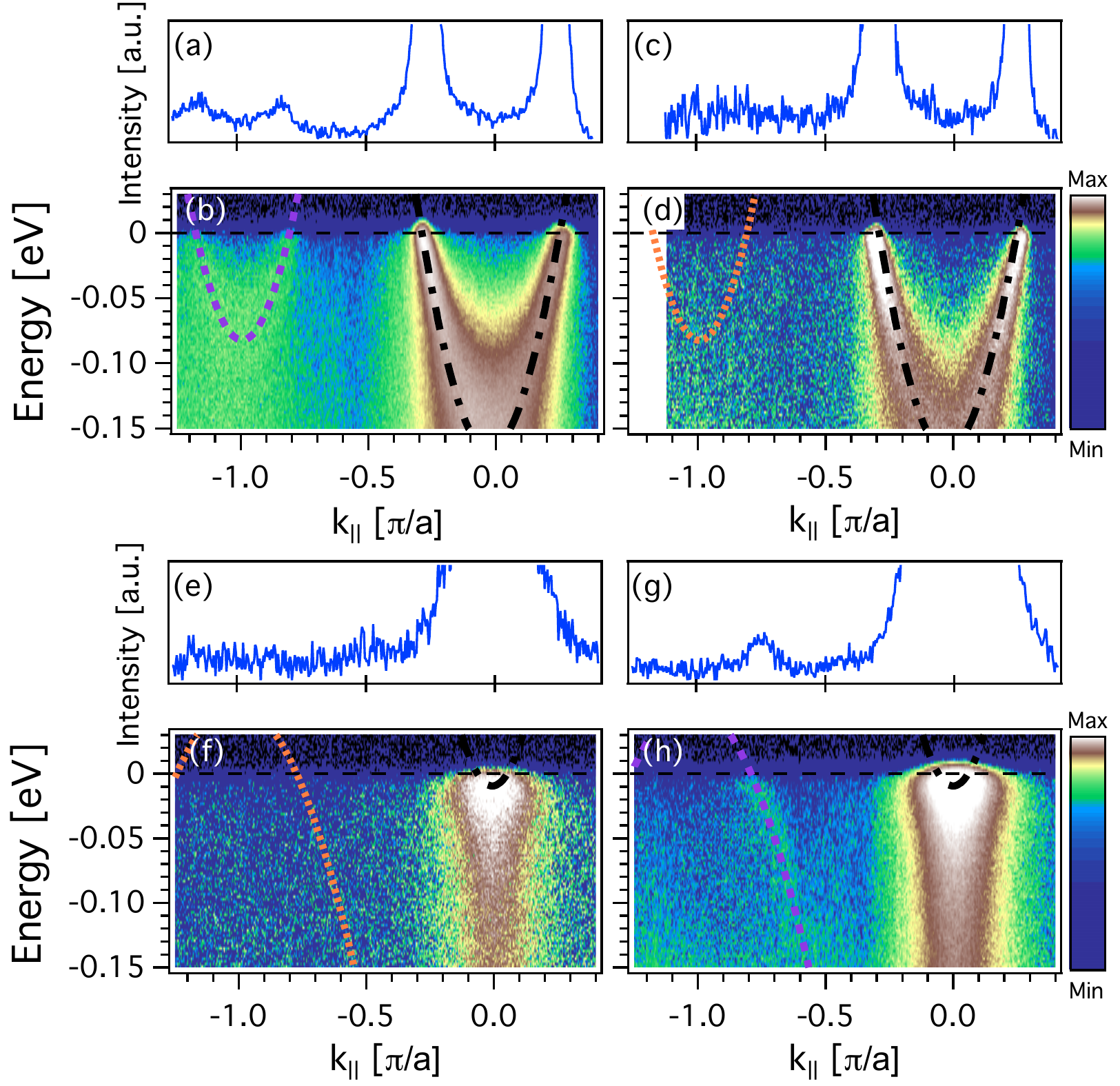}
\caption{ Dispersion of the primary and folded band along selected cuts for La$_{2-x}$Sr$_x$CuO$_4$ ($x = 0.17$). The data were taken at 12 K. (a)-(b) MDC at the $E_F-0.01$ meV  and ARPES intensity map for the cut 1 in \ref{FS_folding17}(c), respectively. (c)-(d) MDC at the $E_F-0.01$ meV  and ARPES intensity map for the cut 2 in \ref{FS_folding17}(d), respectively. (e)-(f) MDC at the $E_F-0.01$ meV  and ARPES intensity map for the cut 3 in \ref{FS_folding17}(c), respectively. (g)-(h) MDC at the $E_F-0.01$ meV  and ARPES intensity map for the cut 4 in \ref{FS_folding17}(d), respectively. The dispersion of the band folded by $\boldsymbol{q}_{\boldsymbol{a}} = (\pi,0)$ (dashed violet line) and $\boldsymbol{q}_{\boldsymbol{b}} = (0, \pi)$ (dotted orange line) are  superimposed to the intensity maps in (b),(h) and  (d),(f), respectively.}\label{Folding_disp}
\end{figure}

To enhance the weak features, in Fig. \ref{FS_folding17}(a) we display the intensity map at $E_F$ in logarithmic  color scale. Besides the primary FS in Map I, in the middle of the intensity plot [see also Fig. \ref{FS_Map_and_LEED} (b)], two weak but clearly visible pieces of FS appear on the left and right sides of the primary FS. In Fig.  \ref{FS_folding17}(b) we plot the $k_F$ extracted from the peak positions in Map I of MDC  at $E_F$ for both the primary FS (red  triangles) and the weak pieces of FS (blue  circles). It can be seen that the two weak pieces of FS mirror the primary FS about the vertical lines at $k_x = \pm \pi /2$, which is equivalent to shifting the primary FS by a commensurate wave vector $\boldsymbol{q}_{\boldsymbol{a}}=(\pi, 0)$ (dashed violet line). The appearance of the weak pieces of FS indicates that a Fermi surface reconstruction related to a wave vector $\boldsymbol{q}_{\boldsymbol{a}}$ occurs in the system. The reconstructed FS is different to the shadow FS previously observed in LSCO \cite{Chang2008, Razzoli2010} and in BSCCO cuprates \cite{Aebi1994} because in those cases the shadow FS is connected to the primary FS by a wave vector along the zone diagonal, i.e. in the $(\pi, \pi)$ direction. It is important to mention that we have observed a reconstructed FS related to the wave vector $\boldsymbol{q}_{\boldsymbol{a}}=(\pi, 0)$, which is nearly  parallel to the cut direction, but found no sign for a reconstruction corresponding to wave vector $\boldsymbol{q}_{\boldsymbol{b}}=(0, \pi)$. The lack of the $\boldsymbol{q}_{\boldsymbol{b}}$-folded band might be due to two different effects;  the $\boldsymbol{q}_{\boldsymbol{b}}$ folded band  is present but not observed  due to ARPES selection rules (the so-called "matrix element effects") \cite{Damascelli2003}  or the this folded band is not present and the $C_4$ rotational symmetry of the system is broken  in favor of  $C_2$ symmetry. To confirm that the lack of $\boldsymbol{q}_{\boldsymbol{b}}$ reconstruction is not due to the matrix element effects in the photoemission process, we rotated the sample about the surface normal (c-axis) by $90^\circ$ and acquired ARPES data with otherwise unchanged experimental conditions. In case of a matrix element effects we would expect to see the band folded again in direction parallel to the (new) cut direction, i.e. along the $(0, \pi)=\boldsymbol{q}_{\boldsymbol{b}}$.  As shown in Map II of Fig. \ref{FS_folding17}(a) and in the corresponding $k_F$ in Fig. \ref{FS_folding17}(b) (light-blue pentagons and green  rhombus),  the reconstructed FS is not displaced along $\boldsymbol{q}_{\boldsymbol{b}}$ but still follow the original folding direction $\boldsymbol{q}_{\boldsymbol{a}}$. This observation demonstrates that the reconstruction of FS is not due to matrix elements effects and it ascertains that the four-fold symmetry is broken in the system. The $\boldsymbol{q}_{\boldsymbol{a}}$ reconstruction is further illustrated in Fig. \ref{Folding_disp} which shows the band dispersions along cut 1-4 as indicated in Fig. \ref{FS_folding17}(b). All the folded-bands associated with the reconstruction can be reproduced after shifting the primary band by wave vector $\boldsymbol{q}_{\boldsymbol{a}}$. On the other hand, no folded-band related to $\boldsymbol{q}_{\boldsymbol{b}}$ reconstruction was observed.

\begin{figure}
\includegraphics[width=0.49\textwidth]{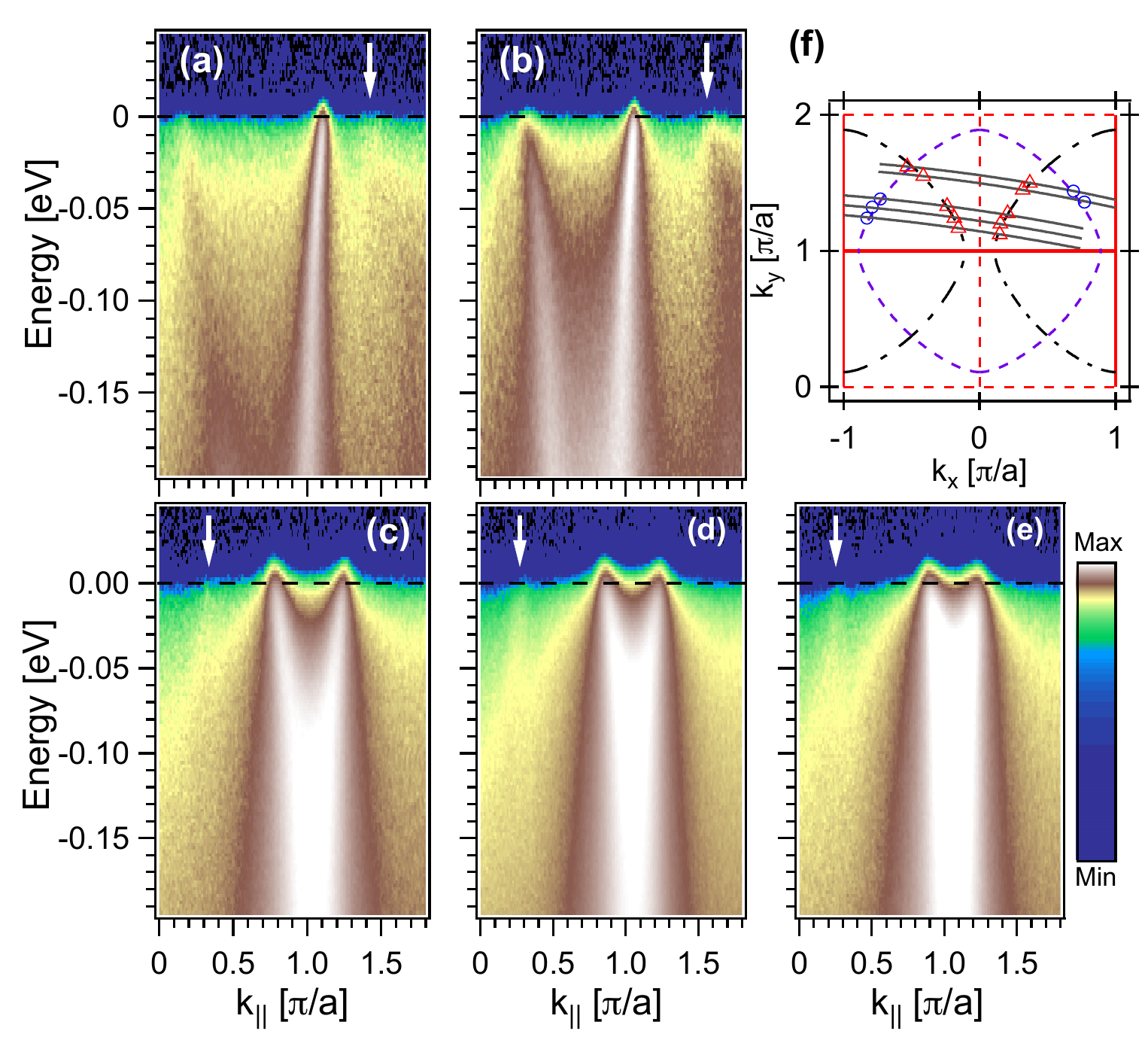}
\caption{ (a)-(e) ARPES Intensity plots of La$_{2-x}$Sr$_x$CuO$_4$ ($x = 0.15$) along the cuts (from top to bottom) shown in (f). The data were taken at 12 K. White arrows indicate the folded band. (f) Black curves are FS of the primary band. The violet curve is a copy of the primary FS, but shifted by $\boldsymbol{q}_{\boldsymbol{a}} = (\pi,0)$. Black lines indicate the momentum cuts. Red triangles and blue empty circles are $k_F$ of the primary and the folded bands, which are determined from MDCs at zero binding energy. }\label{Folding_LSCO15}
\end{figure}

We have investigated the folded bands in LSCO in a wide doping range. For $x \le 0.12$ a folded band related to a shifting of the primary band by $\boldsymbol{q} = (\pi, \pi)$ was observed \cite{Razzoli2010}. For $x \ge 0.22$, except the primary FS and band, there is no indication for any observable reconstructed FS and folded band in our ARPES data acquired in the same experimental conditions. The $\boldsymbol{q}_{\boldsymbol{a}} = (\pi, 0)$ FS reconstruction and the associated band folding appear only in the vicinity of optimally doped samples ($x=0.15, 0.17$). Figure \ref{Folding_LSCO15} shows the ARPES spectra taken from a slightly underdoped LSCO sample with $x = 0.15$. Although the intensity is weaker than the case for $x = 0.17$, the folded bands are still clearly visible, as indicated by the arrows in Figs. \ref{Folding_LSCO15}(a)-(e). The $k_F$ extracted from the peak positions of MDC at $E_F$ show that the folded bands and their FS result from a $\boldsymbol{q}_{\boldsymbol{a}}=(\pi, 0)$  reconstruction [Fig. \ref{Folding_LSCO15}(f)].

%
%
%

%

\section{Discussion}

The $\boldsymbol{q}_{\boldsymbol{a}} = (\pi,0)$  electronic reconstruction in the vicinity of optimal doping of LSCO ($x = 0.15, 0.17$) is different to the previously observed shadow bands and FS duplication along the zone diagonal in cuprates \cite{Chang2008, Razzoli2010, Aebi1994}. This rules out the possibility that the $\boldsymbol{q}_{\boldsymbol{a}} = (\pi,0)$  reconstruction is due to structurally orthorhombic distortions (LTO) of the crystal structure from  tetragonality, because in that case one would expect that a copy of the primary FS is shifted along the $(\pi, \pi)$ direction \cite{Mans2006}. 
A low temperature tetragonal reconstruction (LTT), as the one observed in La$_{2-x}$Ba$_x$CuO$_4$ \cite{Axe1989}, localized at the surface of LSCO could explain the  $(\pi-0)$ reconstruction.
Indeed  LSCO has been reported to be on the verge of a LTO to LTT transition at low temperature \cite{Thurston1989}, which could get stabilized at the surface during the cleaving procedure. 
However while our measurements report that the  $(\pi-0)$ reconstruction is stabilized only in a narrow range of optimal dopings ($x=0.15-0.17$), inelastic neutron-scattering measurements  show that at the LTO-LTT structure instability is stronger at low dopings ($x \le 0.12$)\cite{Kimura2000}. The opposite dependence of the LTO-LTT instability and of the reported $(\pi-0)$
 folding suggests that two effect may not be related. 
The absence in the LEED patterns [see e.g. Fig. 1(a)] of any signature of $1 \times 2$ or $2 \times 1$ surface reconstruction indicates that the folding could result from a dynamic charge modulation or from a nontrivial  structural distortion, similar but different to the case of Bi2212 shadows bands \cite{Mans2006}, where a orthorombic distortion was observed in LEED only at very low energies (below 20 eV) \cite{Strocov2003}. 

The strong doping dependence of the folded bands suggests  that the folding, independently from its origin, is somehow tied to the electronic properties of the sample. 
The $\boldsymbol{q}_{\boldsymbol{a}}$ folded bands were observed at $T = 12$ K which is well below the superconducting temperature 38 K and 35 K for LSCO with $x = 0.15$ and $x = 0.17$, respectively. This suggests that if the $\boldsymbol{q}_{\boldsymbol{a}}$  electronic reconstruction is associated with an instability in the system such as a density wave, the ordering  coexists with superconducting instability.  We note that the $\boldsymbol{q}_{\boldsymbol{a}}$ wave vector associated with the electronic reconstruction is unexpected from the smooth continuation of the incipient incommensurate charge-density-wave along the Cu-O bonding direction, observed in LSCO \cite{Croft2014, Christensen2014, Thampy2014} and other underdoped cuprates \cite{Chang2012, Ghiringhelli2012, Comin2014, SilvaNeto2014}. There, with increasing doping, the incipient incommensurate wave vector is approaching a commensurate one $(\pi/2,0)$, instead of  $\boldsymbol{q}_{\boldsymbol{a}} = (\pi,0)$.  However, one possibility could be that the observed $\boldsymbol{q}_{\boldsymbol{a}}$ corresponds to band-folding associated with two times of  $(\pi/2, 0)$, and the electronic states related to the $(\pi/2, 0)$ reconstruction is too weak to be observed due to the matrix element effects in the photoemission process.
%
Another possibility is that the $\boldsymbol{q}_{\boldsymbol{a}}$ reconstructed FS is related the change of FS topology near $x = 0.17$ \cite{Razzoli2010}. In LSCO, at the doping level slightly above $x = 0.17$, the primary FS changes from a hole-like pocket centered at the $(\pi, \pi)$ point to an electron-like pocket centered at the $\Gamma$ point, the center of the BZ. Accompanying with the topological change of the FS a Van Hove singularity at $(\pi, 0)$ saddle-point approaches the Fermi level. The large and discontinuous density of states near $E_F$ could result in a dynamic charge modulation which leads to a spontaneous breakdown of the point group symmetry \cite{Halboth2000, Gonzalez2001}. 
The charge modulations could involve the whole crystal or the bulk system could be on the verge of a $(\pi, 0)$ electronic reconstruction which get  stabilized only at the surface by the disorder and/or the breaking of translational symmetry, similarly to what was shown for the stripe order in LSCO with $x=1/8$ \cite{Wu2012}. 

Regardless of the exact origin, our observation indicates that at low temperatures an instability associated with the breaking of $C_4$ symmetry coexists with superconductivity near the optimal doping of LSCO. However, it is unclear whether the observed $\boldsymbol{q}_{\boldsymbol{a}} = (\pi, 0)$ electronic reconstruction is general for hole-doped superconducting cuprates, or is particularly related to LSCO.

\section{Conclusions}
In summary, using ARPES we revealed the presence of a weaker folded band which resembles the primary band shifted by  $\boldsymbol{q}_{\boldsymbol{a}} = (\pi, 0)$ in the superconducting state of nearly optimally doped LSCO ($x = 0.15, 0.17$). We show that the absence of a  $\boldsymbol{q}_{\boldsymbol{b}} = (0, \pi)$ folded band is intrinsic but not due to the matrix element effects of the photoemission process, which indicates that the $C_4$ symmetry is broken in the system. The unusual doping dependence of such folded band deserves further study to identify its origin.  

\begin{acknowledgments}
We are grateful to  J. Chang for useful discussions. This work was performed at SLS of the Paul Scherrer Insitut, Villigen PSI, Switzerland and it was supported by the Swiss National Science Foundation through NCCR MaNEP and the Grant No. 200021-137783. E.R. acknowledges support from the Swiss National Science Foundation (SNSF) grant no. P300P2$\_$164649. M.M. and Y.S. acknowledge project funding from the Swedish Research Council (Dnr. 2016-06955). We thank the beam line staff of SIS for their excellent support.
\end{acknowledgments}
%
%
%

\bibstyle{apsrev4-1.bst}

\end{document}